\documentclass[prb,reprint]{revtex4-2}
\usepackage{graphicx}
\usepackage{dcolumn}
\usepackage{bm}
\usepackage[svgnames]{xcolor}
\usepackage{mathtools}
\usepackage{dsfont}
\usepackage{url}
\usepackage{hyperref}
\usepackage[capitalise]{cleveref}
\usepackage{braket}
\usepackage[version=4]{mhchem}
\usepackage{ragged2e}
\usepackage{subcaption}
\usepackage{comment}
\usepackage{amssymb}
\usepackage{mathrsfs}
\usepackage{xcolor}
\usepackage{physics}

\usepackage{ulem}
\usepackage{soul}

\begin{document}

\preprint{APS/123-QED}

\title{Unifying Anderson transitions and topological amplification in non-Hermitian chains}

\author{Cl\'ement Fortin$^1$}
\author{Kai Wang$^{1}$}
\author{T. Pereg-Barnea$^{1,2}$}

\affiliation{$^1$Department of Physics and  Regroupement qu\'{e}b\'{e}cois sur les mat\'{e}riaux de pointe, McGill University, Montr\'eal, Qu\'ebec H3A 2T8, Canada \\
$^2$ICFO-Institut de Ciencies Fotoniques, The Barcelona Institute of
Science and Technology, Castelldefels (Barcelona) 08860, Spain}

\date{\today}

\begin{abstract}


Non-Hermitian systems with non-reciprocal hopping may display the non-Hermitian skin effect, where states under open boundary conditions localize exponentially at one edge of the system. This localization has been linked to spectral winding and topological gain, forming a bulk-boundary correspondence akin to the one relating edge modes to bulk topological invariants in topological insulators and superconductors. In this work, we establish a bulk-boundary correspondence for disordered Hatano-Nelson models.  We relate the localization of states to spectral winding using the Lyapunov exponent and the Thouless formula.  We identify two kinds of phase transitions and relate them to transport properties.  Our framework is relevant to a broad class of 1D non-Hermitian models, opening new directions for disorder-resilient transport and quantum-enhanced sensing in photonic, optomechanical, and superconducting platforms. 
\end{abstract}

\maketitle

\section{Introduction}\label{sec:Intro}
The Hatano-Nelson model \cite{Hatano1996,Hatano1997,Hatano1998} is the prototypical Hamiltonian exhibiting the non-Hermitian skin effect (NHSE) \cite{Yao2018,Martinez2018}, wherein a macroscopic number of eigenstates under open boundary conditions (OBC) are exponentially edge-localized, \textit{i.e.}, skin states. Conversely, the eigenstates under periodic boundary conditions (PBC) are extended and their associated eigenenergies trace a closed curve in the complex plane. The NHSE is a consequence of the following topological correspondence: a nonzero winding number $w(E)$ of the PBC spectral curve about an OBC eigenenergy $E\in\mathds{C}$ guarantees that the associated eigenstate will be localized at a boundary determined by the sign of $w(E)$. 

The winding number about the origin is further tied to topologically-protected directional amplification: $|w(0)|\neq0$ dictates that particles sent into the system pile up exponentially at a boundary in the steady state \cite{Porras2019,Wanjura2020,Fortin2025}. This is reminiscent of the bulk-boundary correspondence in the Su-Schrieffer-Heeger model and can be formulated in terms of the singular value decomposition \cite{Porras2019,Brunelli2023}.

Such non-Hermitian topological statements have long been analytically established for translation-invariant models \cite{Lee2019,Borgnia2020,Okuma2020,Yokomizo2019,Zhang2020,Yang2020,Zhang2022,Brunelli2023} and have been experimentally implemented in diverse platforms, including topoelectrical circuits \cite{Hofman2020,Helbig2020}, coupled optical fiber loops \cite{Weidemann2020}, laser arrays \cite{Liu2022}, ring resonators \cite{Wang2021}, single-photon quantum walks \cite{Xiao2020}, ultra-cold Fermi gas \cite{Zhao2025}, optomechanical networks \cite{Slim2024} and superconducting microwave resonators \cite{Busnaina2024}.

By nature, topological effects are associated with a certain robustness to disorder, which in turn yields interesting and unique effects in non-Hermitian systems \cite{Gong2018,Zhang2020b,Liu2021,Claes2021,Bao2021,Tzortzakakis2021,Longhi2021,Sarkar2022,Liu2023,Zhang2023,Yang2024,Li2024,Longhi2025,Wang2025,Shang2025,Wang2_2025}. However, the index theorem \cite{Okuma2020} and the non-Bloch band theory \cite{Yao2018,Yokomizo2019,Zhang2020,Yang2020} used in clean systems to link spectral winding number and state localization no longer apply to non-Hermitian models when translation-invariance is broken. A new framework is therefore needed to describe disordered non-Hermitian systems.

In this work, we prove that when there is an average asymmetry in hopping amplitudes, the spectral winding number $w(E)$ around an energy $E\in\mathds{C}$ satisfies 
\begin{align}
    \label{eq:BBC}
    w(E) &= \begin{dcases}
        -1 & \Re{L(E)}>0 \\
        0 & \Re{L(E)}<0,
    \end{dcases}
\end{align}
where $L(E) \sim \frac{1}{n}\log \psi_n(E)$ is the Lyapunov exponent defined in the large-$n$ limit for the state 
$\psi(E)=(\psi_n(E))_n$ of energy $E$ obeying the system's dynamical equation. We prove \cref{eq:BBC} in \cref{sec:BBC} for the Hatano-Nelson model subject to disordered on-site potentials and disordered hopping amplitudes. Since $L(E)$ quantifies localization, it predicts when disorder drives a state of energy $E$ from a skin state ($\Re{L(E)}>0$) to an Anderson-localized state ($\Re{L(E)}<0$). We refer to states that are localized by the disorder and decay exponentially to the right as Anderson localized. Energies $E$ with $\Re{L(E)}=0$ are associated with extended state for which the winding number is undefined. We define the mobility edge curve as the set of such energies for which $\Re{L(E)}=0$, corresponding to the PBC spectral curve.

\begin{figure}[ht!]
    \centering
    \begin{subfigure}[b]{\linewidth}
         \centering
         \includegraphics[width=0.78\textwidth]{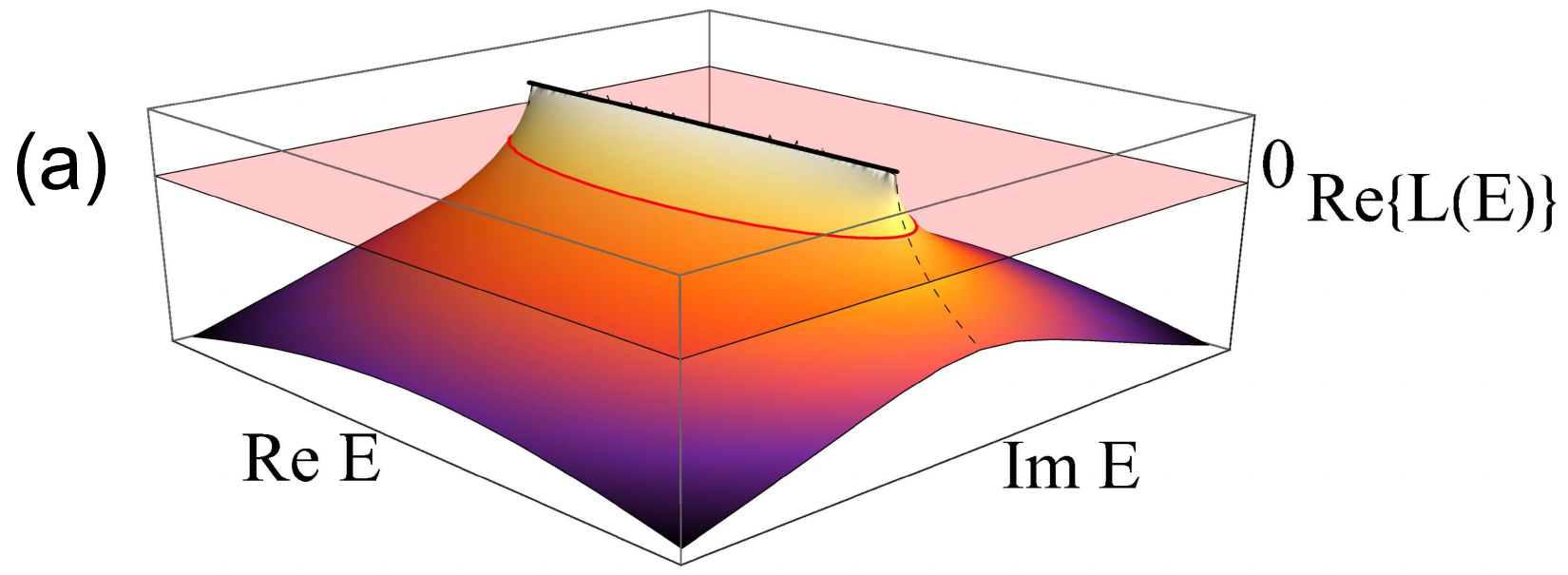}
     \end{subfigure} 
     \\ \vspace{-0.1cm}
      \begin{subfigure}[b]{\linewidth}
         \centering
         \includegraphics[width=0.78\textwidth]{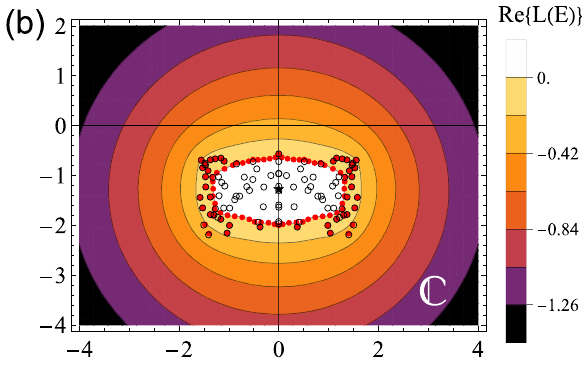}
     \end{subfigure}
    \caption{\justifying (a) Surface plot of the real part of the Lyapunov exponent $\Re{L(E)}$ given by the Thouless formula in \cref{eq:Thouless} as a function of $E\in\mathds{C}$ for the clean Hatano-Nelson model with $J_R=1$ and $J_L=0.5$. The pink plane corresponds to $\Re{L(E)}=0$ and intersects the surface at the red curve (PBC eigenenergies). The black line segment corresponds to the OBC eigenenergies. (b) Contour plot of $\Re{L(E)}$ in the complex plane for the disordered Hatano-Nelson model. The black circles correspond to OBC eigenenergies and the red point to PBC eigenenergies. The PBC eigenvalues that have $\Re{L(E)}=0$ delineate a part of the complex plane that has $\Re{L(E)}>0$ (skin states) with one that has $\Re{L(E)}<0$ (Anderson-localized states). The black star indicates the average OBC eigenenergy. The Hamiltonians are generated by sampling uniformly from $J_{R,n}\in[0.7,1.3]$, $J_{L,n}\in[0,2,0.8]$ and $V_n=-ir_n$ with $r_n\in[0,e]$ and $e=2.718\dotso$ for a chain of 100 sites.}
    \label{fig:BBC}
\end{figure}

\Cref{eq:BBC} makes explicit that the localization of a state $\psi_n(E)\sim e^{L(E)n}$ with energy $E$ is determined by a bulk topological invariant. In particular, it shows that the bulk-boundary correspondence extends to any disordered Hatano-Nelson model, where an invariant computed from bulk properties predicts the existence of boundary-localized states. When all OBC eigenvalues $E$ have \mbox{$w(E)=-1$} ($w(E)=0$), all corresponding eigenstates are skin (Anderson-localized) states. In \cref{sec:Phases}, we use this dichotomy to define two non-Hermitian topological phases: the skin phase and the Anderson-localized phase. Separating these regimes is the mixed phase, in which some OBC eigenstates have a nonzero winding number and others do not. We showcase \cref{eq:BBC} in \cref{fig:BBC} by plotting the Lyapunov landscape of the clean Hatano-Nelson model in \cref{fig:BBC}(a) and the contours of $\Re{L(E)}$ along with the spectra of a fully disordered Hatano-Nelson model in \cref{fig:BBC}(b).

The remainder of this article is organized as follows. In \cref{sec:Phases}, we unify non-Hermitian Anderson localization with topologically-protected directional amplification. In \cref{sec:Thouless}, we use the Thouless formula to define a topological invariant that indicates whether the point gap is open or closed in disordered systems. We then provide detailed results in the case of unidirectional hoppings in \cref{sec:uni}. 

\section{Spectral Winding Number and Lyapunov Exponent}
\label{sec:BBC}
The disordered Hatano-Nelson model is defined by
\begin{align}
    \label{eq:HN}
    \hat{H} &= \sum_n \left(J_{R,n}\hat{a}_{n}^\dagger \hat{a}_{n-1} + J_{L,n}\hat{a}_n^\dagger \hat{a}_{n+1} + V_n\hat{a}_n^\dagger \hat{a}_n\right)
\end{align}
where $J_{R,n},J_{L,n},V_n\in\mathds{C}$. The dynamical equation for a state $\psi=\psi(E)$ of energy $E$ is
\begin{align}
    \label{eq:eom_state}
    J_{L,n}\psi_{n+1} + J_{R,n}\psi_{n-1} &= (E-V_n)\psi_n.
\end{align}
In the rest of the article, we assume a large system size $N\to\infty$ and that the average logarithmic hopping is greater to the right than to the left, \textit{i.e.}, $\mathds{E}\log|J_{R,n}|>\mathds{E}\log |J_{L,n}|$, where we denote by $\mathds{E}(X)$ the expected value of the random variable $X$. We thus expect that $\Re{L(E)}>0$ for a skin state localized on the right edge with energy $E\in\mathds{C}$ and that $\Re{L(E)}<0$ for an Anderson-localized state pinned by disorder, when $n$ is to the right of the pinning site, \textit{i.e.}, for $n$ large enough. When there is no disorder, the PBC spectrum $E(k)=J_Re^{ik}+J_Le^{-ik}$ corresponds to an ellipse in the complex plane with $a=|J_R|+|J_L|$ and $b=|J_R|-|J_L|$. The OBC spectrum $E_n = 2\sqrt{|J_RJ_L|}\cos(\frac{n\pi}{N+1})$ lies inside the ellipse on the real axis, with all the eigenstates skin-localized, that is $\Re{L(E_n)}>0$.

\subsection{Spectral Winding Number}
Establishing \cref{eq:BBC} requires defining a winding number $w(E)\in\mathds{Z}$ that is appropriate for systems subject to disorder. We employ the quantity proposed by Gong \textit{et al}. in Ref.~\cite{Gong2018}, defined by integrating the logarithmic derivative of the non-Hermitian ring's determinant over an artificial flux $\Phi$ threading the ring:
\begin{align}
    \label{eq:winding_num}
    w(E) &= \int_0^{2\pi}\frac{d\Phi}{2\pi i}\, \frac{d}{d\Phi}\log \det (H_\text{pbc}(\Phi)-E).
\end{align}
The introduction of an external flux to compute the spectral winding number in a disordered system is reminiscent of the definition of the TKNN invariants in the integer quantum Hall effect \cite{Klitzing1980,Thouless1982,Avron1983}.
The physical intuition behind \cref{eq:winding_num} can be drawn from an analogy: a changing flux $\Phi$ induces an electromotive force in the periodic system. Specifically, a charged particle whose wavefunction is delocalized should be sensitive to changes in $\Phi$, leading to transport in the ring. It can be shown that not only $H_\text{pbc}(\Phi)=H_\text{pbc}(\Phi+2\pi)$\cite{Gong2018} but also the spectrum of $H_\text{pbc}(\Phi+2\pi/N)$ is the same as that of $H_\text{pbc}(\Phi)$ therefore any flux which is an integer multiple of $2\pi/N$ can be gauged away. Therefore, changing the flux makes delocalized states flow along a curve which interpolates between all delocalized states. This curve traced by this spectral flow in the complex plane is called the mobility edge curve. In contrast, both the energy and wavefunction of an Anderson-localized state should be independent of $\Phi$. For a system with solely Anderson-localized states, the spectral trajectory of $H_\text{pbc}(\Phi)$ does not form a loop (there is no spectral flow) and the winding number around any $E\in\mathds{C}$ is zero. 

In the absence of disorder, the winding number $w(E)$ reduces to 
\begin{align}
    w(E) = \int_0^{2\pi} \frac{dk}{2\pi i} \dv{}{k} \log \det (H_\text{pbc}(k)-E)
\end{align}
where $k$ is a quasimomentum label. By block diagonalizing $\det H_\text{pbc}(\Phi) = \prod_k \det H_\text{pbc}(k+\Phi/N)$, then $\partial_\Phi \log \det H_\text{pbc}(\Phi) = \frac{1}{N}\sum_k\partial_k \log\det H_\text{pbc}(k+\Phi/N)$. In other words, the external flux and the momentum label play the same role, except that varying $\Phi$ from $0$ to $2\pi$ makes each PBC eigenvalue flow to the next one, while varying $k$ from $0$ to $2\pi$ makes each eigenvalue trace the entire PBC spectral curve until it comes back to itself.

\subsection{Derivation of the Winding Number in Terms of the Lyapunov Exponent}
\label{subsec:deriv}
The Hamiltonian matrix for a periodic chain of $N$ sites threaded by an external flux $\Phi$ is
\begin{align}
     H_\text{pbc}^{(N)}(\Phi) &= \begin{pmatrix}
        V_1 & J_{L,1} & 0 & \dots & J_{R,1}e^{-i\Phi} \\
        J_{R,2} & V_2 & J_{L,2} & \dots & 0 \\
        0 & J_{R,3} & V_3 & \ddots & \vdots \\
        \vdots & \vdots &\ddots & \ddots & J_{L,N-1} \\
        J_{L,N}e^{i\Phi} & 0 & \dots & J_{R,N} & V_{N}
    \end{pmatrix},
\end{align}
where the gauge (which places the flux on the 1-L links) is chosen without loss of generality. We may relate the PBC characteristic polynomial, denoted by $p_{N}^\text{pbc}(E,\Phi)\equiv \det (H_\text{pbc}^{(N)}(\Phi)-E)$, with that of the OBC via
\begin{align}
    \begin{split}
    p_{N}^\text{pbc}(E,\Phi) &= p_{1,N}^\text{obc}(E) - J_{L,N}J_{R,1}p_{2,N-1}^\text{obc}(E) \\
    &\quad- (-1)^N\left[e^{i\Phi}\prod_{n=1}^N J_{L,n} + e^{-i\Phi} \prod_{n=1}^N J_{R,n}\right],
    \end{split}
\end{align}
where $p_{i,j}^\text{obc}(E)$ is the characteristic polynomial of the Hamiltonian from site $i$ to site $j$ subject to open boundary conditions, for which there is no flux. Replacing $e^{i\Phi}$ by $z$, the argument principle \cite{Krantz1999} gives that the winding number around $E\in\mathds{C}$ is $w(E) = Z-P$ where $Z$ and $P$ are the numbers of zeroes and poles of $f_N(z_N) = p_{N}^\text{pbc}(E,z_N)$ respectively. There is a pole at $z_N=0$ for all $N$. For the zeroes, we can rearrange the expression to 
\begin{align}
    \begin{split}
        \label{eq:derivation_step}
        1 &= z_N\left(\frac{p_{1,N}^\text{obc}(E)}{\prod_{n=1}^N (-J_{R,n})}- \frac{J_{L,N}}{J_{R,N}}\frac{p_{2,N-1}^\text{obc}(E)}{\prod_{n=2}^{N-1}(-J_{R,n})}\right)\\
        &\qquad -z_N^2\prod_{n=1}^N \frac{J_{L,n}}{J_{R,n}}.
    \end{split}
\end{align}
We now want to express $p_{i,j}^\text{obc}(E)$ in terms of the Lyapunov exponent $L(E)$ to relate the number of zeroes in the unit circle to the sign of $\Re{L(E)}$. To do so, we use the retarded Green's function $G_{n,m}(t) \equiv i\Theta(t)\expval{[\hat{a}_n(t),\hat{a}^\dagger_m(0)]}$. In frequency space, it's equation of motion is
\begin{align}
    \label{eq:eom_G}
    \begin{split}
    J_{L,n}G_{n+1,m}(E) &+ J_{R,n}G_{n-1,m}(E) \\
    &= (E-V_n)G_{n,m}(E)+\delta_{n,m},
    \end{split}
\end{align}
see \cref{apdx:Geom} for details. For $n\neq m$, this is the same equation as \cref{eq:eom_state} for the states, so we find $G_{n,m}(E) \sim  e^{L(E)(n-m)}$. In particular, we can write $G_{N,1}(E) = -\frac{A}{J_{R,1}}e^{L(E)N}$ for some constant $A\in\mathds{C}$. From \cref{eq:eom_G}, we have $G(E) = (H_\text{obc}^{(N)}-E)^{-1}$, or equivalently
\begin{align}
    G_{N,1}(E) &= \frac{1}{p_{1,N}^\text{obc}(E)}[\text{adj}(H_\text{obc}^{(N)}-E)]_{N,1}
\end{align}
where the adjugate matrix in this case is 
\begin{align}
    [\text{adj}(H_\text{obc}^{(N)}-E)]_{N,1} = \prod_{n=2}^N (-J_{R,n})
\end{align}
due to the tridiagonal structure of $H_\text{obc}^{(N)}$. Thus,
\begin{align}
    \label{eq:sec2_Thouless}
    \lim_{N\to\infty} \frac{1}{N} \log p_{1,N}^\text{obc}(E) &= \mathds{E}\log (-J_{R,n}) - L(E).
\end{align}
We remark that the Lyapunov exponent only depends on the energy $E\in\mathds{C}$ and is independent of $n,N$. Substituting this expression in \cref{eq:derivation_step}, the zeroes $z_N$ of the function $f_N$ in the limit $N\to\infty$ then have to satisfy
\begin{widetext}
\begin{align}
    0 &= \lim_{N\to\infty}\frac{1}{N}\log \left|\frac{e^{-L(E)N}}{A} \left(1-\frac{J_{L,N}}{J_{R,N}}e^{2L(E)}\right)z_N-z_N^2\prod_{n=1}^N \frac{J_{L,n}}{J_{R,n}}\right|.
\end{align}
\end{widetext}
In the large-$N$ limit, the law of large numbers gives
\begin{align}
    \prod_{n=1}^N \frac{J_{L,n}}{J_{R,n}} &\sim \exp\left(N\mathds{E}\log\frac{J_{L,n}}{J_{R,n}}\right).
\end{align}
Since we assumed $\mathds{E}(\log |J_{R,n}|) > \mathds{E}(\log|J_{L,n}|)$, the above term goes to zero exponentially fast as $N\to\infty$. We remark that if the term $z_N^2\prod_{n=1}^N\frac{J_{L,n}}{J_{R,n}}$ does not go to zero exponentially fast, then $|z_N|$ goes to infinity exponentially fast and hence does not contribute to the winding number (since we only care about those zeroes that are within the unit circle). Otherwise, we are left with
\begin{align}
    \label{eq:derivation_step2}
    0 &= \lim_{N\to\infty} \frac{1}{N}\log \left|\frac{e^{-L(E)N}}{A}\left(1-\frac{J_{L,N}}{J_{R,N}}e^{2L(E)}\right)z_N\right|.
\end{align}
Assuming $ L(E) \neq \frac{1}{2}\log \frac{J_{R,N}}{J_{L,N}}$, then 
\begin{align}
    \lim_{N\to\infty}\frac{1}{N}\log \left|\frac{1}{A}\left(1-\frac{J_{L,N}}{J_{R,N}}e^{2L(E)}\right)\right| = 0
\end{align}
and \cref{eq:derivation_step2} reduces to 
\begin{align}
    \Re{L(E)} &= \lim_{N\to\infty} \frac{1}{N}\log |z_N|,
\end{align}
or $|z_N|\sim e^{\Re{L(E)}N}$. We hence find that the winding number around any point $E\in\mathds{C}$ is given by \cref{eq:BBC}.

If a skin state becomes Anderson-localized by disorder, \cref{eq:BBC} dictates that the associated energy $E$ must escape the area enclosed by the mobility edge curve.
Equivalently, the mobility edge curve must adjust so as to no longer enclose the point $E$, with the transition happening when $E$ lies on the curve. When the OBC eigenenergy $E$ escapes out of the (closed) mobility edge curve so that $w(E)=0$, it is accompanied by a PBC eigenenergy with approximately the same value, see \cref{fig:BBC} (b). Physically, this is because an Anderson-localized state does not ``sense" the system's boundaries. Therefore, the Anderson-localized eigenstates and eigenenergies under OBC and PBC are the same to a good approximation. Since the total number of states under OBC and PBC is also the same, each OBC skin state of energy $E$ with $w(E)=-1$ comes with a PBC extended state of energy $E'$ with $\Re{L(E')}=0$. Thus, there is a region with $w(E)=-1$ in the complex plane even when the system only possesses one skin state. 

At the skin-to-Anderson-localized transition of a particular OBC eigenstate $\psi_n(E)\sim e^{L(E)n}$ of energy $E$, the correlation length diverges, $\Re{L(E)}=0$. The associated state $\psi_n(E)$ and the spectral Green's function $G_{n,1}(E)$ then have a power-law scaling in $n$, which is typical of correlations at transitions. Such power-law scalings have previously been investigated using the large deviation principle \cite{Huang2025}. 

\begin{figure}[t!]
    \centering
    \includegraphics[width=0.78\linewidth]{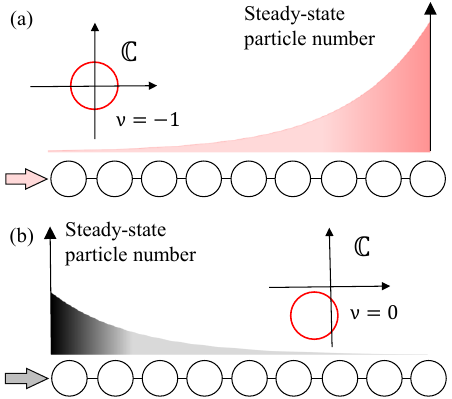}
    \caption{\justifying Schematic of the non-Hermitian topology behind directional amplification discussed in \cref{sec:Phases}. Depending on the value of the spectral winding number around the origin $\nu$ there may be (a) amplification or (b) decay of particle number in the steady state.}
    \label{fig:steady_state_topology}
\end{figure}

\begin{table*}[ht!]
    \centering
    \begin{minipage}[t]{0.48\textwidth}
    \vspace{0pt}
    \centering
    \begin{tabular}{||c|c||} 
     \hline
     Winding number & Topological Phase  \\ 
     \hline\hline
     $\nu=-1$ & Amplification \\ 
     \hline
     $\nu=0$ & Decay \\
    \hline
     $w_\Gamma=-1$ & Skin or mixed (point gap open)\\
     \hline
     $w_\Gamma=0$ & Anderson-localized (point gap closed) \\ 
     \hline
    \end{tabular}
    \caption{\justifying The spectral winding numbers corresponding to different non-Hermitian topological phases. The winding number around the origin $\nu\equiv w(0)$ is responsible for the phase of amplification or decay in the steady state (long-time limit), see \cref{sec:Phases}. The winding number $w_\Gamma\equiv w(\mathds{E}(V_n))$ indicates whether the point gap is open or closed, see \cref{sec:Thouless}.}
    \label{tab:Phases}
    \end{minipage}
    \hfill
    \begin{minipage}[t]{0.44\textwidth}
        \vspace{0pt}
        \centering
        \includegraphics[width=\linewidth]{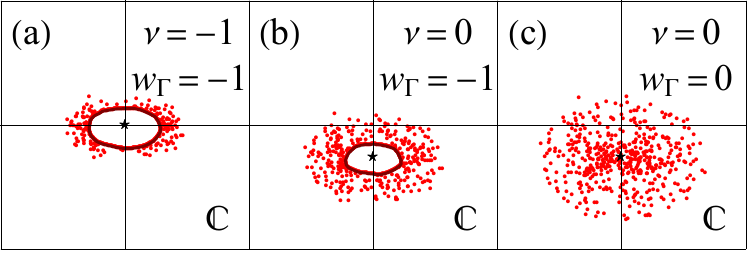}
        \captionof{figure}{\justifying The PBC spectrum (red dots) can either lie on the mobility edge curve (black) or be an isolated point, depending on disorder strength. The mean on-site potential value $\mathds{E}(V_n)$ (depicted by a black star) is the point at which the real part of the Lyapunov exponent is greatest for a single-band model in the large-chain limit.}
        \label{fig:Phases}
    \end{minipage}
\end{table*}

\section{Classification of Topological Phases}
\label{sec:Phases}
The correspondence between the spectral winding number $w(E)$ and the Lyapunov exponent $L(E)$ of the associated state allows us to define three distinct phases for disordered Hatano-Nelson models. 

(I) \textit{Skin phase}: if $\Re{L(E)}>0$ for all OBC eigenenergies, every eigenstate is a skin state localized at the (right) boundary. In terms of the Lyapunov exponent, the skin phase is characterized by all of the OBC eigenenergies $E$ satisfying 
\begin{align}
    0 < \Re{L(E)} \leq \frac{1}{2}\bigg(\mathds{E}\log|J_{R,n}| - \mathds{E}\log|J_{L,n}|\bigg).
\end{align}
We show this in \cref{apdx:skin_phase}. To find a condition on $J_{L,n},J_{R,n}$ and $V_n$ to ensure the system remains in the skin phase, we demand that $\Re{L(E)}>0$ for all OBC eigenenergies $E$. A sufficient requirement for this is
\begin{align}
    \label{eq:skin_phase_condition}
    |E-V_n| \leq |J_{R,n}| -|J_{L,n}|
\end{align}
for all $n=1,\dots,N$. In other words, the relative detuning and/or loss $|E-V_n|$ on each site $n$ needs to be less than the relative hopping amplitudes $|J_{R,n}|-|J_{L,n}|$, see \cref{apdx:skin_phase} for the proof. We note that the condition in \cref{eq:skin_phase_condition} is by no means necessary for a skin phase. 

(II) \textit{Mixed phase}: if some OBC eigenstates have $\Re{L(E)}>0$ and others have $\Re{L(E)}<0$, skin and Anderson-localized states coexist. This is exemplified in \cref{fig:BBC} (b), where some OBC eigenenergies lie outside of the mobility edge curve, corresponding to skin states, while some lie inside, corresponding to Anderson-localized states pinned by disorder.

(III) \textit{Anderson-localized phase}: if $\Re{L(E)}<0$ for all OBC eigenenergies, every eigenstate is exponentially localized in the bulk, pinned by disorder. In this case, the mobility edge curve no longer exists as there are no more extended PBC eigenstates. 

Starting from the skin phase and increasing disorder, the OBC skin states and PBC extended states undergo Anderson localization one by one, making the mobility edge curve shrink. When an OBC eigenstate of energy $E$ Anderson-localizes, it transitions from $\Re{L(E)}>0$ to $\Re{L(E)}<0$. Increasing disorder beyond a critical value closes the point gap ($w(E)=0$ for all $E\in\mathds{C}$) and drives a non-Hermitian Anderson transition analogous to the metal-insulator transition in Hermitian systems \cite{Anderson1958}. At this point, $\Re{L(E)}<0$ for the entire PBC spectrum and the mobility edge curve disappears. We remark that the sign of $\Re{L(E)}$ is a consequence of our choice that $\mathds{E}\log|J_{R,n}|>\mathds{E}\log |J_{L,n}|$.

In addition to the above three phases, the system can simultaneously be in either one of two phases, enabling another type of bulk-boundary correspondence. Since the equation of motion of the Green's function is the same as that of the states, the zero-energy Green's function $G(E=0)$ is tied to the winding number 
\begin{align}
    \nu\equiv w(0).
\end{align}
 The importance of the zero-energy response $G(0)$ stems from its description of scattering experiments in the steady state when the input fields are coherent drives in the system's rotating frequency, see \cref{apdx:SS}. In particular, $\nu=-1$ ($\Re{L(0)}>0$) corresponds to \textit{topological amplification}: particles injected at a site $n$ pile up exponentially on the right boundary. In contrast, $\nu=0$ ($\Re{L(0)}<0$) corresponds to exponential decay, with particles exponentially localized at the input site or the pinning site of the zero-energy mode, see \cref{fig:steady_state_topology}. The topological phase transition at $\Re{L(0)}=0$ delineates the phase of exponential amplification from that of decay. The presence of amplification depends on whether the eigenstates can coherently carry the input field across the chain at long times. The existence of skin states is hence not a sufficient condition for amplification, since we may still have $\nu=0$. On the other hand, one skin state is sufficient for the point gap to be open. If it is such that the mobility edge curve winds around the origin, then this skin state alone is enough to give rise to exponential amplification.
 
Our Lyapunov-exponent approach therefore provides a unifying framework for understanding non-Hermitian Anderson transitions and topological amplification, allowing us to predict localization properties and transport behavior across a broad class of one-dimensional models. Moreover, our approach encompasses earlier formulations of topological amplification based on singular value decomposition \cite{Porras2019}, Dyson's equation \cite{Wanjura2020,Wanjura2021,Brunelli2023}, and the residue theorem \cite{Zirnstein2021a,Zirnstein2021b,Xue2021}. See \cref{tab:Phases,fig:Phases}.

Furthermore, such non-Hermitian topology is the basis for non-Hermitian quantum-enhanced sensing protocols \cite{McDonald2020,Koch2022,Bao2023,Parto2025}. Understanding how disorder affects this topology is thus crucial to designing quantum sensors that are robust to noise and defects.


\section{Thouless Formula and the Point Gap}
\label{sec:Thouless}

As shown by \cref{eq:sec2_Thouless}, the Lyapunov exponent $L(E)$ can be expressed in terms of the determinant of the Hamiltonian matrix under OBC. Using the law of large numbers,
\begin{align}
    \begin{split}
    \lim_{N\to\infty}\frac{1}{N}\log |p_{1,N}^\text{obc}(E)| &= \lim_{N\to\infty}\frac{1}{N} \log \prod_{n=1}^N |E_n-E| \\
    &= \mathds{E}(\log|E_n-E|)
    \end{split}
\end{align}
and so, defining $\log J_R \equiv \mathds{E}\log |J_{R,n}|$, \cref{eq:sec2_Thouless} yields
\begin{align}
    \label{eq:Thouless}
    \Re{L(E)} &= -\mathds{E}\left(\log\left|\frac{E_n-E}{J_R}\right|\right).
\end{align}
This is the so-called Thouless formula \cite{Thouless1972} and it is defined on the whole complex plane. It provides a way of computing the real Lyapunov exponent landscape without working with large products of transfer matrices. For example, we may readily compute the Lyapunov exponent of each OBC eigenstate when there is no disorder, see \cref{apdx:ReL_clean}. We note that $\Re{L(E)}$ may not diverge when evaluated at OBC eigenvalues $E=E_n$.

In the disordered Hatano-Nelson model, we have shown in \cref{sec:BBC} that the sign of the real part of the Lyapunov exponent gives the winding number of the mobility edge curve around the point $E$. Additionally, we use the Lyapunov exponent to extend the notion of a point gap size to disordered systems. To this end, we assume that the system consists of a single point gap (single-band system). Nonetheless, the framework can be readily generalizable to systems with multiple point gaps.

In the context of spectral topology, it is often useful to define the point gap, which characterizes the energy spectrum of a non-Hermitian Hamiltonian \cite{Kawabata2019}. It is customary to define the size of the point gap $\Gamma$ in translation-invariant non-Hermitian systems by the smallest distance between the mobility edge curve and its geometric center. Since non-Hermitian systems often come with symmetries, the geometric center of the mobility edge curve is usually readily identifiable. In finite and disordered systems, the size of the point gap may become ambiguous when the spectral curve lacks a clear geometric center. To remedy this, we choose the center to be the point $E_c\in\mathds{C}$ at which the real part of the Lyapunov exponent $\Re{L(E_c)}$ is the greatest. This represents the complex energy whose associated state is most strongly skin-localized on the right edge. As disorder is increased and the system enters its Anderson-localized phase, the point gap thus closes.  This happens at $E_c\in\mathds{C}$ when $\Re{L(E_c)}=0$. 

We now argue that  $E_c=\mathds{E}(V_n)$, so that all $E\in\mathds{C}$ satisfy $\Re{L(\mathds{E}(V_n))}\gtrsim \Re{L(E)}$. First, the fact that $\Re{L(\mathds{E}(E_n))}\gtrsim \Re{L(E)}$ for all $E$ is intuitively understood from the fact that the mean $\mathds{E}(E_n)$ is the minimizer of $E\mapsto \mathds{E}|E_n-E|^2$ and should thus roughly minimize $\mathds{E}\log|E_n-E|$. By the Thouless formula of \cref{eq:Thouless}, $\Re{L(E)} = \mathds{E}\log|J_{R,n}|-\mathds{E}\log|E_n-E|$. Therefore, the mean $\mathds{E}(E_n)$ should roughly maximize $\Re{L(E)}$. Second, since $\tr(H_0+V) = \tr(H_0) + \tr(V)$ for $H_0$ the Hamiltonian matrix of the clean Hatano-Nelson model and $V$ the disorder matrix, then 
\begin{align}
    \label{eq:means}
    \mathds{E}(E_n) &= \mathds{E}(V_n)
\end{align}
for $E_n$ the eigenenergies of $H_0+V$. This is true since the average eigenvalue of $H_0$ is zero. Note that \cref{eq:means} holds regardless of boundary conditions. We conclude that $\Re{L(\mathds{E}(V_n))}\gtrsim \Re{L(E)}$ for all $E\in\mathds{C}$, see \cref{apdx:PointGap} for a specific example. Therefore, we define the size of the point gap by
\begin{align}
    \label{eq:point_gap}
    \Gamma \equiv \min_{E\in \{\Re{L(E)}=0\}} |\mathds{E}(V_n)-E|,
\end{align}
where the minimum is over all points on the mobility edge curve. Consequently, when the system undergoes a transition from a mixed phase to an Anderson-localized phase, the point gap closes at $\mathds{E}(V_n)$. Therefore, the winding number 
\begin{align}
    w_\Gamma \equiv w(\mathds{E}(V_n))
\end{align}
is the topological invariant indicating whether the point gap is open or closed: it is open when $\omega_\Gamma=-1$ and is closed when $\omega_\Gamma=0$; see \cref{tab:Phases} and \cref{fig:Phases}. 

The Thouless formula also allows us to determine the Lyapunov exponent far from the domain of OBC eigenvalues. Indeed, when $|\mathds{E}(E_n)-E|\gg |\mathds{E}(E_n)-E_n|$ for all OBC eigenvalues $E_n\in\mathds{C}$, then 
\begin{align}
    \begin{split}
    \mathds{E}\log|E_n-E| &= \mathds{E}\log|E_n-\mathds{E}(E_n)+\mathds{E}(E_n)-E| \\
    &\approx \log|\mathds{E}(E_n)-E|.
    \end{split}
\end{align}
Since $\mathds{E}(E_n) = \mathds{E}(V_n)$ then
\begin{align}
    \label{eq:ReL_continuation}
    \Re{L(E)} \approx \log(\frac{J_R}{|\mathds{E}(V_n)-E|}).
\end{align}
In particular, driving the system far out of resonance yields a response that decays according to
\begin{align}
    G_{j,1}(E) &\approx \left(\frac{J_R}{|\mathds{E}(V_n)-E|}\right)^{j-1}
\end{align}
in the large-$j$ limit, and that is independent of the $J_{L,n}$. See \cref{apdx:PointGap} for a specific example.

When the on-site potentials are not disordered, say $V_n=0$, the point gap is open ($w_\Gamma=-1$) as long as $\mathds{E}\log |J_{R,n}|> \mathds{E}\log|J_{L,n}|$ with 
\begin{align}
    \Re{L(0)} &= \frac{1}{2}\Big(\mathds{E}\log |J_{R,n}|-\mathds{E}\log |J_{L,n}|\Big)>0,
\end{align}
which is readily obtained from \cref{eq:sec2_Thouless} and the recurrence relation of the characteristic polynomial 
\begin{align}
    \label{eq:recrelOBC}
    \begin{split}
    p_{1,N}^\text{obc}(E) &= (V_N-E)p_{1,N-1}^\text{obc}(E) \\
    &\quad- J_{R,N}J_{L,N-1}p_{1,N-2}^\text{obc}(E).
    \end{split}
\end{align}
This is what is observed in Ref.~\cite{Claes2021}. Namely, starting with a closed point gap $w_\Gamma=0$ with $J_R=J_L$ and $V_n=0$, we may open the point gap $w_\Gamma\neq 0$ by adding asymmetric disorder to the hopping amplitudes such that $\mathds{E}\log |J_{R,n}|\neq \mathds{E}\log|J_{L,n}|$.

Surprisingly, there is a way to add some disordered on-site potentials without closing the point gap; it suffices that every other on-site potential value be constant. Suppose $V_{2n}=V$ for all $n$ and some $V\in\mathds{C}$, keeping $V_{2n+1}\in\mathds{C}$ and $J_{R,n}, J_{L,n}\in\mathds{C}$ general. If $N=2n$, the recurrence relation of \cref{eq:recrelOBC} evaluated at $E=V$ yields
\begin{align}
    p_{1,2n}^\text{obc}(V) &= \prod_{m=1}^n (-J_{R,2m}J_{L,2m-1}).
\end{align}
If $N=2n+1$, 
\begin{align}
    \begin{split}
    p_{1,2n+1}^\text{obc}(V) &= \sum_{m=0}^n (V_{2m+1}-V) \\
    & \times\prod_{j=m+1}^n (-J_{R,2j+1}J_{L,2j})\prod_{i=1}^{m} (-J_{R,2i}J_{L,2i-1})
    \end{split}
\end{align}
Since $\frac{1}{2n-1}\log\sum_{m=0}^{n-1} (V_{2m+1}-V)\to 0$ by the law of large numbers, \cref{eq:sec2_Thouless} gives
\begin{align}
    \Re{L(V)} &= \frac{1}{2}\Big(\mathds{E}\log |J_{R,n}|-\mathds{E}\log |J_{L,n}|\Big).
\end{align}
Therefore, no matter how disordered the $V_{2n+1}$ are, the point gap remains open with $w_\Gamma=-1$, while $\nu$ may or may not be nonzero. This effect is a generalized result of what is found in Ref.~\cite{Fortin2025}, where it was shown only for $V=0$, such that $\nu=w_\Gamma$. We remark that this robustness is in stark contrast to what happens in Hamiltonians with symmetric hopping, where any finite disorder induces Anderson localization \cite{Anderson1958}.

\begin{figure}[t!]
    \centering
    \includegraphics[width=0.48\textwidth]{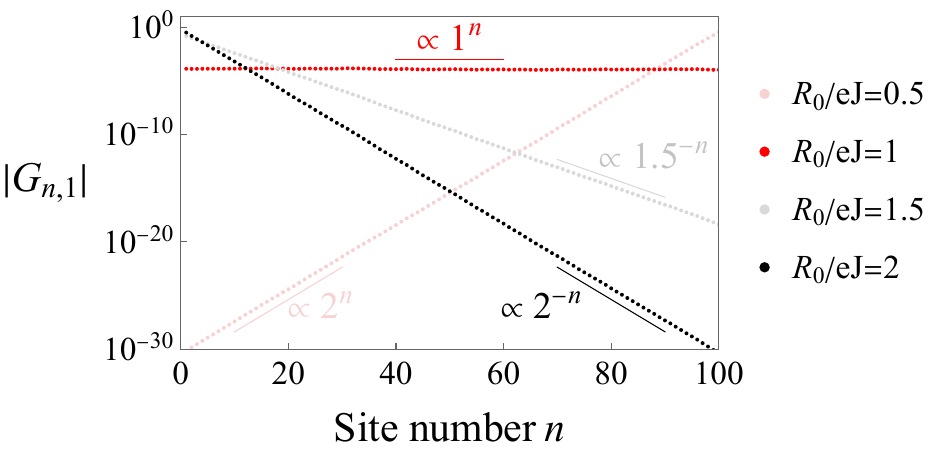}
    \caption{\justifying Normalized absolute value of the Green's function in the steady state ($E=0$) for a unidirectional chain of 100 sites subject to disorder. The dots represent the numerical Green's function averaged over 1000 disorder realizations for on-site potentials $V_n=e^{i\phi_n}r_n$ with uniform radial probability distribution $g(r_n)=1/R_0$ over $r_n\in[0,R_0]$. The data is in excellent agreement with our analytical result \cref{eq:SS_greenf}, as demonstrated by the solid line segments.}
    \label{fig:Gplot}
\end{figure}

\begin{figure*}[ht!]
    \centering
     \begin{subfigure}[b]{0.48\textwidth}
         \centering
         \text{Uniform Phase Distribution, $h(\phi_n)=1/2\pi$}\par
         \includegraphics[width=\textwidth]{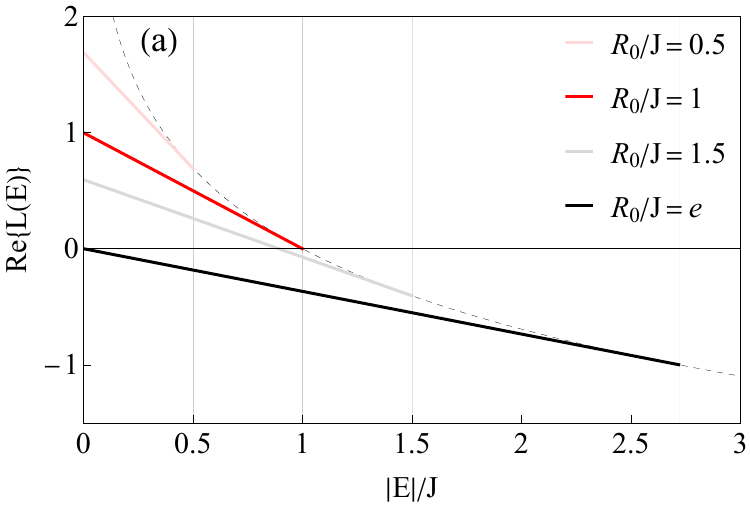}
     \end{subfigure} 
      \begin{subfigure}[b]{0.48\textwidth}
         \centering
         \text{Constant Phase, $h(\phi_n)=\delta(\phi_n-\theta)$}\par
         \includegraphics[width=\textwidth]{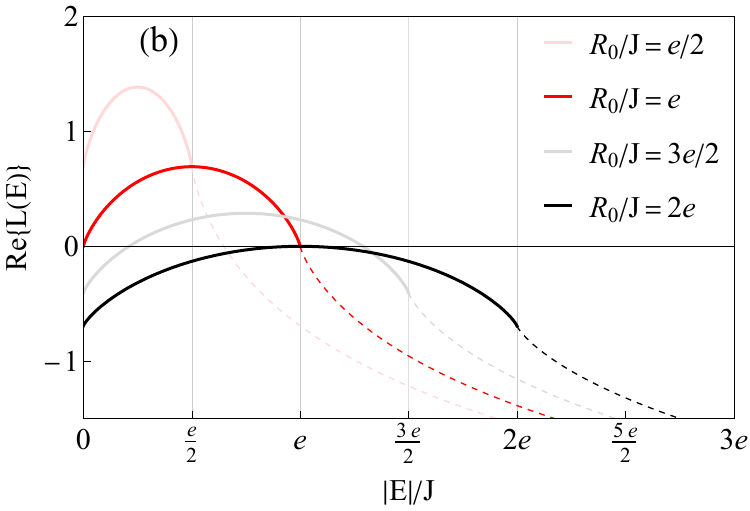}
     \end{subfigure}
    \caption{\justifying Real part of the Lyapunov exponent in terms of the energy $|E|$ for a uniform radial distribution of $V_n$ over $[0,R_0]$. (a) The solid line segments correspond to the curve given by \cref{eq:ReL_uni} obtained with a uniform phase distribution of $V_n$ over the domain of OBC eigenvalues. The dashed line is the curve of \cref{eq:ReL_uni_outside}. When $|E|>R_0$, it is the continuation of $\Re{L(E)}$ outside of the domain of OBC eigenvalues. (b) The solid line segments correspond to the curve given by \cref{eq:ReL_dirac} over the range of OBC eigenvalues obtained when the phase of the on-site potentials $V_n$ is constant, $\phi_n=\theta$. The dashed lines are the continuation of $\Re{L(E)}$ outside of the domain of OBC eigenvalues.}
    \label{fig:ReL}
\end{figure*}

\section{Unidirectional Hopping}
\label{sec:uni}
To obtain more quantitative results, we apply the previous framework to the disordered Hatano-Nelson model with unidirectional hopping amplitudes. Specifically, we assume that $J_{L,n}=0$ for all $n=1,\dots,N$ and that $\{(J_{R,n},V_n)\}_{n=1}^N$ are independent and identically distributed. In this case, the eigenenergies under OBC are $E=V_j$ and the associated eigenstates
\begin{align}
    \psi_n^{(j)}(E) &= \frac{1}{\mathcal{N}}\begin{cases}
        0 & 1\leq n<j, \\
        1 & n=j, \\
        \prod_{m=j+1}^n \frac{J_{R,m}}{E-V_m} & j<n\leq N
    \end{cases}
\end{align}
where $\mathcal{N}$ is the normalizing constant. The Lyapunov exponent for the energy $E\in\mathds{C}$ is
\begin{align}
    L(E) &= \mathds{E}\log(J_{R,n})-\mathds{E}\log(E-V_n).
\end{align}
If $f$ is the distribution of potentials, then
\begin{align}
    \label{eq:uniThouless}
    \Re{L(E)} &= \mathds{E}\log|J_{R,n}| - \int dV_n f(V_n) \log|E-V_n|.
\end{align}
In what follows, we define $\log J \equiv \mathds{E}\log |J_{R,n}|$. Since $H_\text{pbc}^{(N)}(\Phi)$ is lower triangular up to a single entry in the upper right corner, the proof \cref{eq:BBC} is much simpler than the one presented in \cref{sec:BBC}, see \cref{apdx:BBC_uni}.

\subsection{Topological Phase Transition}
\label{sec:uni_SSPT}

We first look at the topological phase transition probed by the spectral winding number $\nu$ when the on-site potentials $V_n=r_ne^{\phi_n}$ are distributed according to $f(V_n)=g(r_n)h(\phi_n)$. For $E=0$, \cref{eq:uniThouless} yields
\begin{align}
    \label{eq:SSint}
    \Re{L(0)} &= \int_0^{\infty} dr_n g(r_n)\int_0^{2\pi} \frac{d\phi_n}{2\pi} h(\phi_n) \log(\frac{J}{|r_ne^{i\phi_n}|}).
\end{align}
The topological phase transition happens when 
\begin{align}
    \label{eq:SSPT}
    \int_0^\infty dr_n\, g(r_n) \log(\frac{J}{r_n})&= 0.
\end{align}
In particular, the above condition is independent of the phase distribution $h(\phi_n)$. This independence was identified in Appendix B.3 of Ref.~\cite{Gong2018}. In the context of topological amplification, it means that the steady-state response of the OBC system is only tied to the localization of the zero-energy state, regardless of the amount of eigenstates that are Anderson-localized. For instance, when the radial distribution is uniform over $[0,R_0]$, that is $g(r_n)=1/R_0$, \cref{eq:SSint} gives $\Re{L(0)} = \log(eJ/R_0)$ and the steady-state Green's function is
\begin{align}
    \label{eq:SS_greenf}
    |G_{n,1}(0)|\sim \left(\frac{eJ}{R_0}\right)^n,
\end{align}
where $e=2.718\dotso$ is Euler's number. We plot this theoretical prediction with the numerical value of the Green's function in \cref{fig:Gplot} and find excellent agreement. 

In \cref{apdx:periodic_pot}, we generalize this results to Hatano-Nelson chains made up of unit cells of length $L$ with different potential distributions. 

In the following two subsections, we show the effect of the phase distribution on the localization of eigenstates by studying two extreme cases: (i) when the phase is uniformly distributed, $h(\phi_n)=1/2\pi$, and (ii) when the phase is constant, $h(\phi_n)=\delta(\phi_n-\theta)$, with $\theta\in[0,2\pi)$. While the topological phase transition is independent of $h(\phi_n)$, the point gap closure is not. We plot the real part of the Lyapunov exponent in each case in \cref{fig:ReL}.

\subsection{Uniform Phase Distribution, $h(\phi_n)=1/2\pi$}

The case of uniform phase distribution with constant unidirectional hopping amplitudes $J_{R,n}=J$ has already been studied in Ref.~\cite{Longhi2021}, and generalizing the results to disordered hopping amplitudes is straightforward. Nonetheless, we apply our framework for completeness. 

When the phase distribution is uniform, the average potential value is $\mathds{E}(V_n)=0$. Since the OBC eigenenergies are rotationally-invariant, the mobility edge curve (when it exists) forms a circle, and we know from \cref{sec:Thouless} that $\Re{L(0)}\geq \Re{L(E)}$ for all $E\in\mathds{C}$. In particular, the point gap closes at the origin of the complex plane when the topological phase transition given by \cref{eq:SSPT} is satisfied. That is, $\nu=w_\Gamma$. Particles sent into the system from site 1 will hence tend to pile-up exponentially on site $N$ as long as the point gap is still open. Moreover, the system remains in the skin phase as long as $R_0<J$ with the point gap $\Gamma$ of the mobility edge curve satisfying $\Gamma=J$, which was found not to depend on the form of $g$ in Ref.~\cite{Longhi2021}. For \mbox{$|E|\geq R_0$}, using \cref{eq:ReL_continuation} with $\mathds{E}(V_n)=0$ shows that
\begin{align}
    \label{eq:ReL_uni_outside}
    \Re{L(E)} &= \log(\frac{J}{|E|}).
\end{align} 
For $|E|\leq R_0$, we proceed by assuming that the radial distribution $g(r_n)=1/R_0$ is uniform on $[0,R_0]$. In that case, \cref{eq:uniThouless} gives
\begin{align}
    \label{eq:ReL_uni}
    \Re{L(E)} &= \log (\frac{eJ}{R_0}) - \frac{|E|}{R_0}
\end{align}
with steady-state topological amplification as long as $ R_0<eJ$. We plot the Lyapunov exponent in \mbox{\cref{fig:ReL} (a)} on the domain of the OBC energies, $|E|\in[0,R_0]$. We also plot the energy spectra in \cref{fig:ReL_spectra}. If $R_0>J$, setting $\Re{L(E)}=0$ in \cref{eq:ReL_uni_outside,eq:ReL_uni} and solving for $\Gamma=|E|$ gives the size of the point gap. All in all, 
\begin{align}
    \label{eq:unipointgap}
    \Gamma &= \begin{dcases} 
        J & R_0\leq J, \\
        R_0\log (\frac{eJ}{R_0}) & R_0\geq J.
    \end{dcases}
\end{align}

\noindent \Cref{tab:UniPhases} summarizes at what disorder value the two winding numbers of \cref{tab:Phases} change. We plot the point gap $\Gamma/J$ as a function of $R_0/J$ in \cref{fig:pointgap}. We note that some data points lie further than one error bar away from our theoretical prediction. This is simply explained by the fact that we are interested in computing the point gap, which is here defined as the minimum distance between the mobility edge curve and  $\mathds{E}(V_n)\approx 0$, see \cref{eq:point_gap}. Since our theoretical point gap in \cref{eq:unipointgap} is the radius of the mobility edge curve, it does not take into account that the curve drawn by the numerically computed PBC eigenvalues may exhibit small deviations from a circle. We observe that this discrepancy becomes more subtle when disorder is increased since variations between different disorder realizations become larger.

\subsection{Constant Phase, $h(\phi_n)=\delta(\phi_n-\theta)$}
\begin{figure*}[ht!]
    \centering
    \begin{subfigure}{\textwidth}
        \includegraphics[width=\textwidth]{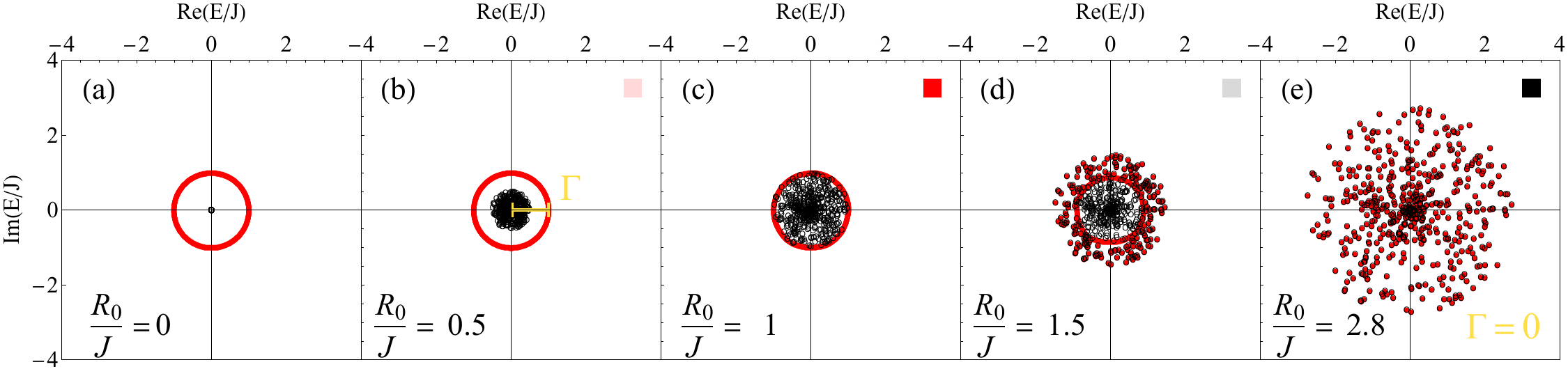}
    \end{subfigure}
    \\
    \begin{subfigure}{\textwidth}
        \includegraphics[width=\textwidth]{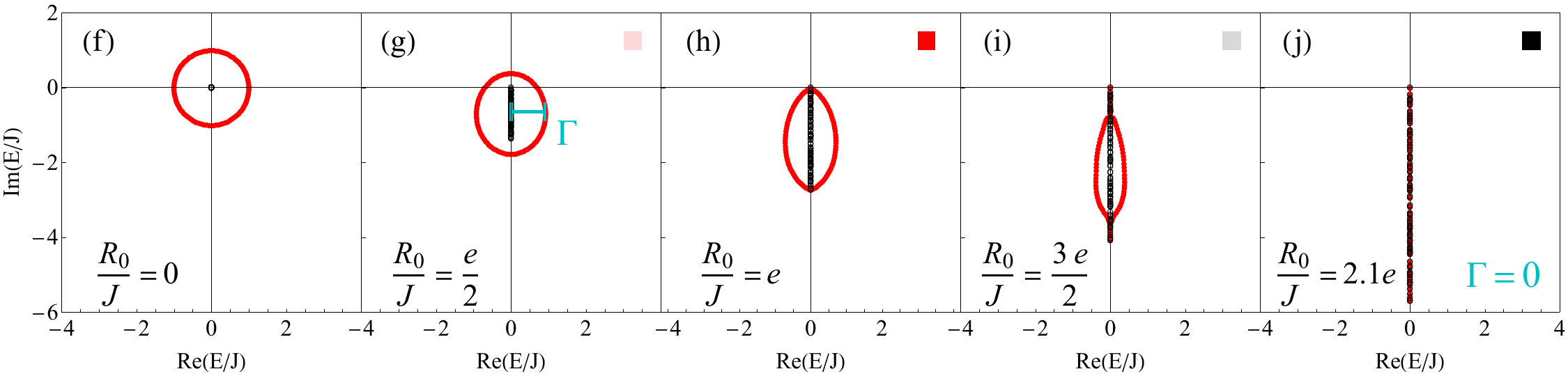}
    \end{subfigure}
    \caption{\justifying The OBC (black circles) and PBC (red dots) spectrum of the unidirectional model with uniform on-site potential radial distribution $g(r_n)=1/R_0$ over $r_n\in[0,R_0]$. (a)--(e) The on-site potential phase distribution is uniform, $h(\phi_n)=1/2\pi$. The topological phase transition and the closure of the point gap $\Gamma$ happen at the same time when $R_0=eJ$, see (e). (a)--(c) The skin phase corresponds to $R_0\in[0,J)$, (d) the mixed phase corresponds to $R_0\in(J,eJ)$ and (e) the Anderson-localized phase to $R_0\in(eJ,\infty)$. The plots correspond to a chain of 500 sites. (f)--(j) The phase distribution is a Dirac delta function $h(\phi_n)=\delta(\phi_n-\theta)$ with $\theta=-\pi/2$. The topological phase transition happens when $R_0=eJ$, see (h), while the point gap $\Gamma$ closes when $R_0=2eJ$, see (j). (f)--(h) The skin phase corresponds to $R_0\in[0,eJ)$, (i) the mixed phase to $R_0\in(eJ,2eJ)$ and (j) the Anderson-localized phase to $R_0\in(2eJ,\infty)$. The plots correspond to a chain with 100 sites. The colored squares on the top right corner of the plots identify the color of the associated curve in \cref{fig:ReL}.}
    \label{fig:ReL_spectra}
\end{figure*}

When the potentials all share the same phase $\theta$, that is $V_n=r_ne^{i\theta}$, for some $r_n>0$, the OBC spectrum $E_n=V_n$ lies on a line and 
\begin{align}
    \label{eq:ReL_uni_constant_phase}
    \Re{L(E)} &= \int_0^{\infty} dr_n\,g(r_n)\log(\frac{J}{|Ee^{-i\theta}-r_n|}).
\end{align}
Writing $E=(x + iy) e^{i\theta}$, we find that
\begin{align}
    \Re{L((x+iy)e^{i\theta})} &= \log J - \frac{1}{2}\mathds{E}\log ((x-r_n)^2+y^2)  \\
    &\leq \Re{L(xe^{i\theta})}.
\end{align}
In other words, the Lyapunov exponent is greatest on the line $\{xe^{i\theta}:x\in\mathds{R}\}$ in the complex plane whose phase matches that of the on-site potentials. 
Assuming that the radial distribution $g(r_n)$ is uniform on $[0,R_0]$, the real part of the Lyapunov exponent on this line is
\begin{align}
    \begin{split}
    \label{eq:ReL_dirac}
    \Re{L(xe^{i\theta})} &= \log(\frac{eJ}{R_0}) \\
    &+ \frac{x}{R_0}\log\left|\frac{R_0}{x}-1\right| - \log\left|1-\frac{x}{R_0}\right|.
    \end{split}
\end{align}

\begin{table}[t!]
    \begin{tabular}{||c|c|c||} 
     \hline
     $w(E)\backslash h(\phi_n)$ & $1/2\pi$ & $\delta(\phi_n-\theta)$ \\ 
     \hline
     $\nu$ & $R_0/eJ=1$ & $R_0/eJ=1 $\\ 
     \hline
     $w_\Gamma$ & $R_0/eJ=1$ & $R_0/eJ=2$ \\ 
     \hline
    \end{tabular}
    \caption{\justifying The steady-state and point-gap winding numbers $\nu$ and $w_\Gamma$ change at different values of radial disorder $R_0$ depending on $h$. The topological phase transition is independent of $h$ and happens at $R_0/eJ=1$, but this is not the case for the point-gap phase transition, see \cref{fig:pointgap}.}
    \label{tab:UniPhases}
\end{table}

While the topological phase transition again happens when $R_0=eJ$, the point gap here closes when $R_0 = 2eJ$ at the value $E=(R_0/2)e^{i\theta}$. The zero energy OBC eigenstate is thus the first to undergo an edge-to-Anderson-localized transition. In other words, disorder in the potentials induces a topological phase transition although all OBC eigenstates are still edge-localized. To understand why, it suffices to look at when the maximum value of the curve \cref{eq:ReL_dirac} is zero, see \cref{apdx:ReL_dirac_curve}. We plot the real part of the Lyapunov exponent in \mbox{\cref{fig:ReL} (b)} along the curve $E=|E|e^{i\theta}$ for $|E|\geq0$. We also plot the energy spectra in \cref{fig:ReL_spectra} for $\theta=-\pi/2$. For this choice of phase, the potentials hence correspond to on-site loss terms. 

\begin{figure}[t!]
    \centering
    \includegraphics[width=0.45\textwidth]{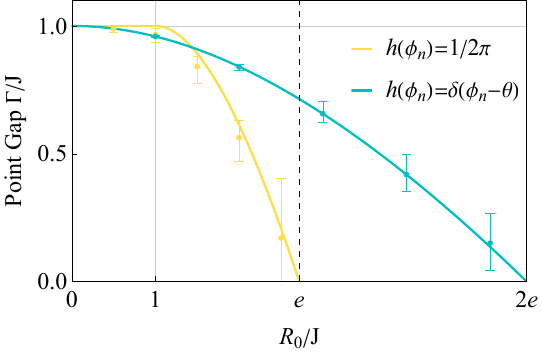}
    \caption{\justifying The point gap $\Gamma/J$ as a function of disorder strength $R_0/J$ for the uniform radial distribution $g(r_n)=1/R_0$ on $r_n\in[0,R_0]$. Yellow data corresponds to a uniform phase distribution and cyan data to potentials with a constant phase. The yellow solid line is given by \cref{eq:unipointgap} and the cyan solid line is obtained by solving \cref{eq:diracpointgap}. The error bars are the maximum and minimum values of the point gap over 50 realizations of disorder for a chain of 500 sites. The vertical dashed line at $R_0/J=e$ represents the disorder strength at which the system undergoes a topological phase transition, which is independent of the phase distribution, see \cref{sec:uni_SSPT}. }
    \label{fig:pointgap}
\end{figure}

Contrary to the case of uniform phase distribution, the mobility edge curve is no longer circular but ellipse-like. Using the framework of \cref{sec:Thouless}, the center is the point $E$ maximizing $\Re{L(E)}$, namely $E=(R_0/2)e^{i\theta}$. To characterize the point gap, we note that on average the minimum distance between the mobility edge curve and the center happens for points $E=(x+iy)e^{i\theta}$ on the minor axis, with $x=R_0/2$. To find the value of the point gap $\Gamma=y$, we set $\Re{L((x+i\Gamma)e^{i\theta})}=0$ and perform the average in \cref{eq:ReL_uni_constant_phase} to find
\begin{align}
    \label{eq:diracpointgap}
    0 &= \log(\frac{eJ}{\sqrt{\Gamma^2+R_0^2/4}}) - \frac{2\Gamma}{R_0}\arctan(\frac{R_0}{2\Gamma}).
\end{align}
See \cref{tab:UniPhases} for a summary of the different disorder-induced transitions. We plot the point gap $\Gamma/J$ against $R_0/J$ in \cref{fig:pointgap} and find excellent agreement between the numerical data and our theoretical prediction. The point gap is numerically computed by finding the closest point on the mobility edge curve to the average on-site potential $\mathds{E}(V_n)$, as per \cref{eq:point_gap}. We observe that it is also possible to find the height of the ellipse-like shape traced by the mobility edge curve by setting \cref{eq:ReL_dirac} equal to zero and solving for $x\in[0,R_0]$.

\section{Discussion}
Our results thus far apply to both fermionic and bosonic particle systems. However, they can further be extended to bosonic quadratic Hamiltonians of the form
\begin{align}
    \begin{split}
    \hat{H}_{BKC} &= \sum_n\left( it_n\hat{a}^\dagger_{n+1}\hat{a}_n + i\Delta_n \hat{a}^\dagger_{n+1}\hat{a}_n^\dagger +\text{H.c.}\right)  \\
    &\qquad-\sum_{n} \frac{i\gamma_n}{2} \hat{a}_n^\dagger \hat{a}_n
    \end{split}
\end{align}
where $t_n$ is a hopping amplitude, $\Delta_n>0$ is a two-mode squeezing amplitude and $\gamma_n>0$ ($\gamma_n<0$) accounts for a loss (gain) rate. We assume $t_n>\Delta_n>0$. This Hamiltonian corresponds to the bosonic Kitaev chain \cite{McDonald2018}. In the quadrature basis $\hat{x}_n = (\hat{a}_n + \hat{a}_n^\dagger)/2$ and $\hat{p}_n = (\hat{a}_n-\hat{a}_n^\dagger)/2i$, the Heisenberg-Langevin equation of motions are
\begin{subequations}
\begin{align}
    \dot{\hat{x}}_n &= (t_n+\Delta_n)\hat{x}_{n-1} - (t_n-\Delta_n)\hat{x}_{n+1} - \frac{\gamma_n}{2}\hat{x}_n \\
    &= -i\sum_{m=1}^N (M_x)_{n,m}\hat{x}_{m},
\end{align}
\end{subequations}
\begin{subequations}
\begin{align}
    \dot{\hat{p}}_n &=(t_n-\Delta_n)\hat{p}_{n-1} - (t_n+\Delta_n)\hat{p}_{n+1} - \frac{\gamma_n}{2}\hat{p}_{n} \\
    &= -i\sum_{m=1}^N (M_p)_{n,m}\hat{p}_{m}
\end{align}
\end{subequations}
where $M_x$ and $M_p$ are dynamical matrix blocks of the form given in \cref{eq:HN}. We observe that the dynamics along the chains are unidirectional when $t_n=\Delta_n$ for all $n$. The quadrature-quadrature Green's functions 
\begin{align}
    G^{(q)}(E) &\equiv (M_q-E)^{-1}, \quad q=x,p
\end{align}
satisfy $G^{(p)}_{n,m}(E) = (-1)^{n+m}G^{(x)}_{m,n}(E)$. Therefore, the quadrature winding numbers have opposite sign: $w^{(x)}(E) = -w^{(p)}(E)$. The framework and results outlined above hence apply both to $M_x$ and $M_p$, with opposite edges identified. 

In this work, we showed that the Lyapunov exponent $L(E)$ characterizes the localization of any state $\psi(E)$ with energy $E$ that satisfies the system's dynamical equation. In the context of non-Hermitian systems, the strength of this quantity is that it is directly linked to the spectral winding number $w(E)$ for any one-dimensional disordered system with complex-valued nearest-neighbor hopping and complex-valued on-site potentials. Furthermore, since the Green's function satisfies the same dynamical equation as the states, the Lyapunov exponent also provides topological information about scattering experiments and gives an illuminating perspective on topologically-protected directional amplification in disordered systems. Using the determinant representation of the Lyapunov exponent (the Thouless formula), we define a topological invariant that dictates whether the point gap is closed or open. Moreover, we derive many analytical results for general unidirectional systems. We highlight the role of the potential's phase distribution in unidirectional systems by studying the topological transitions of two extreme cases. Specifically, when the potentials follow a uniform phase distribution over $[0,2\pi]$ and when they have a constant phase, such as when potentials represent loss rates. 

Our results can readily be applied to other well-known 1D systems, particularly those with symmetries, such as non-Hermitian versions of the SSH model \cite{Schomerus2013,Yao2018} and the bosonic Kitaev chain \cite{McDonald2018} subject to on-site loss. Studying the connections between our Lyapunov exponent framework and recent amoeba formulations of non-Bloch band theory in higher dimensions \cite{Wang2024} would also be a particularly interesting avenue of research. 

Moreover, our framework naturally lends itself to experimental realizations, where disorder and noise are ubiquitous. A significant advantage of experimentally investigating the localization of states in a disordered system is that one does not have to implement both PBC and OBC; disorder alone can dictate a state's localization, which can be probed via its energy. It is thus unnecessary to experimentally realize boundaries to study different topological regimes, which may be beneficial to synthetic dimensions implementations.

\textit{Note Added}--While finalizing this paper, we became aware of a related preprint reporting similar results \cite{Sun2025}. 

\begin{acknowledgements}
    We thank Maciej Lewenstein for useful discussions. We acknowledge the support from the Québec's Ministère de l'Économie, de l'Innovation et de l'Énergie (MEIE), Photonique Quantique Québec (PQ2), Natural Sciences and Engineering Research Council of Canada (NSERC) [ALLRP 588334-23], the Fonds de recherche du Québec (FRQNT) and Regroupement québécois sur les matériaux de pointe (RQMP).
\end{acknowledgements}

\appendix 
\crefalias{section}{appendix}
\section{Frequency Space Green's Function}
\label{apdx:Geom}
In this appendix, we derive the equation of motion of the retarded Green's function in frequency space. We start from the definition
\begin{align}
    G_{n,m}(t) &\equiv i\Theta(t)\expval{[\hat{a}_n(t),\hat{a}_m^\dagger(0)]}.
\end{align}
Taking the time derivative and using $\partial_t\Theta(t) = \delta(t)$,
\begin{align}
    \begin{split}
    i\partial_t G_{n,m}(t) &= -\delta(t)\expval{[\hat{a}_n(t),\hat{a}_m^\dagger(0)]} \\
    &\qquad +i\Theta(t)\expval{[i\partial_t \hat{a}_n(t),\hat{a}_m^\dagger(0)]},
    \end{split}
\end{align}
where commutation relations at equal times are $[\hat{a}_n,\hat{a}^\dagger_m]= \delta_{n,m}$. Heisenberg's equation of motion yields 
\begin{align}
    i\partial_t \hat{a}_n &= -[H_\text{obc}^{(N)},\hat{a}_n] \\
    &= J_{R,n}\hat{a}_{n-1} + J_{L,n}\hat{a}_{n+1} + V_n\hat{a}_n.
\end{align}
Therefore, 
\begin{align}
    \begin{split}
    i\partial_t G_{n,m}(t) &= -\delta(t) \delta_{n,m} + J_{R,n} G_{n-1,m}(t) \\
    &\qquad + J_{L,n}G_{n+1,m}(t) +V_nG_{n,m}(t).
    \end{split}
\end{align}
Taking the Fourier transform gives $i\partial_t \to E$ and $\delta(t)\to 1$. Rearranging gives the frequency space equation of motion of \cref{eq:eom_G}. From this equation of motion, we also see that 
\begin{align}
    (H_\text{obc}^{(N)}-E)G(E) = \mathds{1},
\end{align}
or that $G(E) = (H_\text{obc}^{(N)}-E)^{-1}$.

\section{Skin Phase Estimates}
\label{apdx:skin_phase}
We here provide proofs on two claims made in \cref{sec:Phases} regarding the skin phase. First, we prove that any OBC eigenstate of energy $E$ that is a skin state has a Lyapunov exponent satisfying  
\begin{align}
    \label{eq:skinphase}
    0<\Re{L(E)} \leq \frac{1}{2}\Big(\mathds{E}\log |J_{R,n}| - \mathds{E}\log |J_{L,n}|\Big).
\end{align}
The lower bound follows by definition for a skin state, so we only have to show the upper bound. To this end, consider the transfer matrix in the \textit{left} direction
\begin{align}
    \begin{pmatrix}
        \psi_{n} \\
        \psi_{n-1} 
    \end{pmatrix}
    &= T_{n} \begin{pmatrix}
        \psi_{n+1} \\
        \psi_{n}
    \end{pmatrix}, \quad T_n \equiv \begin{pmatrix}
        \frac{E-V_n}{J_{R,n}} & -\frac{J_{L,n}}{J_{R,n}} \\
        1 & 0
    \end{pmatrix}
\end{align}
whose determinant is $|\det T_n| = |J_{L,n}/J_{R,n}|$. The real part of the Lyapunov exponent has
\begin{align}
    \Re{L(E)} &= -\lim_{N\to\infty}\frac{1}{N}\mathds{E}\log \norm{\prod_{n=N-1}^{2} T_n},
\end{align}
for any matrix norm $\norm{\cdot}$, where the matrix product is defined by $\prod_{n=N-1}^{2} T_n \equiv T_2T_3\dots T_{N-1}$. The minus sign arises because the transfer matrix is defined in the direction of decreasing $n$, opposite our definition of the Lyapunov exponent, $\psi_n(E) \sim e^{L(E)n}$. This choice is crucial for the upper bound, which follows from the inequality $\log\norm{A} \geq \frac{1}{2}\log|\det A|$ with 
\begin{align}
    \left|\det\left(\prod_{n=N-1}^2 T_n\right)\right| &= \prod_{n=N-1}^{2} |\det T_n| = \prod_{n=N-1}^{2} \left|\frac{J_{L,n}}{J_{R,n}}\right|
\end{align}
and the law of large numbers. We stress that using the transfer matrix in the opposite direction to the Lyapunov exponent is essential for obtaining the upper bound. 
The second claim is that $|E-V_n|\leq |J_{R,n}|-|J_{L,n}|$ for all $n=1,\dots,N$ is a sufficient condition to ensure that the system is in the skin phase. To show this, note that the submultiplicity of the norm in the transfer matrix representation of $\Re{L(E)}$ yields
\begin{align}
    \begin{split}
    \Re\{{L}(E)\} &= -\lim_{N\to\infty}\frac{1}{N}\mathds{E}\log\norm{\prod_{n=N-1}^2 {T}_n}\\
    &\geq -\mathds{E}\log \norm{T_n}.
    \end{split}
\end{align}
Since this is valid for any matrix norm, we may use the $\infty$-norm to obtain
\begin{align}
    \log \norm{T_n}_\infty &= \log\max\left\{\left|\frac{E-V_n}{J_{R,n}}\right|+ \left|\frac{J_{L,n}}{J_{R,n}}\right|,1\right\}.
\end{align}
We hence find that $\Re\{L(E)\} >0$ as long as
\begin{align}
    \label{eq:skin_phase_bound}
    |E-V_n|\leq |J_{R,n}| - |J_{L,n}|
\end{align}
for all $n=1,\dots,N$. 

\section{The Zero-Energy Green's Function Describes the Steady-State Response}
\label{apdx:SS}
Suppose that each site $n$ of the system is subject to coherent driving $\beta_n\in\mathds{C}$ at frequency $\omega_d$ with coupling constant $\kappa_n>0$. The equation of motion of the field $\hat{a}_n$ obtained from input-output theory \cite{Clerk2010} is
\begin{align}
    \dot{\hat{a}}_n(t) &= -i\sum_{m=1}^NH_{n,m} \hat{a}_m(t) -\sqrt{\kappa_n}\beta_n e^{-i\omega_d t},
\end{align}
where $H_{n,m}$ is the effective Hamiltonian matrix under OBC. Using the ansatz $\hat{a}(t) = \hat{A} e^{-i\omega_d t}$ for the steady-state, 
\begin{align}
    -i\omega_d \hat{A}_n &= -i\sum_{m=1}^N H_{n,m}\hat{A}_m -\sqrt{\kappa_n}\beta_n.
\end{align}
Rearranging, we find 
\begin{align}
    \hat{A}_n &= i\sum_{m=1}^N [(H-\omega_d)^{-1}]_{n,m}\sqrt{\kappa_m}\beta_m
\end{align}
Since $[(H-\omega_d)^{-1}]_{n,m}=G_{n,m}(\omega_d)\sim -e^{L(\omega_d)(n-m)}$ and $\expval{\hat{a}_n^\dagger \hat{a}_n} = \expval{\hat{A}_n^\dagger \hat{A}_n}$, then the average particle number on site $n$ in the steady state is 
\begin{align}
    \expval{\hat{a}_n^\dagger \hat{a}_n}^\text{ss} &\sim \sum_{m=1}^N\kappa_m |\beta_m|^2 e^{2\Re{L(\omega_d)}(n-m)}.
\end{align}
When $\Re{L(\omega_d)}>0$, coherent particles incident upon site $m$ will thus be exponentially localized on the right edge (site $N$). On the other hand, they will be exponentially localized in the bulk, pinned by disorder, when $\Re{L(\omega_d)}<0$. We note that it is customary to absorb the driving frequency into the Hamiltonian by adding detuning, so that $\omega_d=0$. In this case, it is the winding number $\nu$, given by the sign of $\Re{L(0)}$, that dictates whether or not there is amplification. 

\section{Lyapunov Exponent for the Clean Hatano-Nelson Model}
\label{apdx:ReL_clean}
We here compute the real part of the Lyapunov exponent for each OBC eigenvalue of the clean Hatano-Nelson model using the Thouless formula introduced in \cref{sec:Thouless}. When there is no disorder, \textit{i.e.}, when $V_n=0$ and $|J_{R,n}|=|J_R|>|J_L|=|J_{L,n}|$ for all $n$, all OBC eigenstates are skin states localized on the right edge whose Lyapunov exponent saturate the upper bound in \cref{eq:skinphase}. In particular, the spectrum under OBC is $E_n = 2\tilde{J}e^{i\theta}\cos(n\pi/(N+1))$ for $n=1,\dots,N$ where $\tilde{J}e^{i\theta}=\sqrt{J_RJ_L}$ for real $\tilde{J}$. The OBC spectrum hence lies on the line of angle $\theta$. For any $E=(x+iy)e^{i\theta}$ we thus have 
\begin{align}
    \begin{split}
    \Re{L(x+iy)e^{i\theta}} &= \frac{1}{2}\mathds{E}\log\left(\frac{|J_R|^2}{(x-E_ne^{-i\theta})^2+y^2}\right)\\
    &\leq \Re{L(xe^{i\theta})}.
    \end{split}
\end{align}
In other words, the real part of the Lyapunov exponent is greatest on the line of angle $\theta$ (dashed line in \mbox{\cref{fig:BBC} (a)} where $\theta=0$). To obtain the value of $\Re{L(xe^{i\theta})}$, we start from the Thouless formula of \cref{eq:Thouless}:
\begin{align}
    \begin{split}
    \Re{L(xe^{i\theta})} &= \log \left|J_R\right|\\
    &\quad -\int_0^1 dt \log \left|x-2\tilde{J}\cos(\pi t)\right|.
    \end{split}
\end{align}
Writing
\begin{align}
    &\int_0^1 dt \log \left|x-2\tilde{J}\cos(\pi t)\right| \\
    &\qquad= \log|x|+ \int_0^\pi \frac{dt}{2\pi} \log (1-a\cos(t))^2 \label{eq:int_table}
\end{align}
where $a=2\tilde{J}/x$, the integral in \cref{eq:int_table} can be found in a table of integrals (\textit{e.g.}, BI (330)(1) in Ref.~\cite{GR8}):
\begin{align}
    \int_0^\pi &\frac{dt}{2\pi} \log (1-a\cos(t))^2 \\
    &\qquad= \begin{dcases}
        \frac{1}{2}\log \frac{a^2}{4} & |a|<1, \\
        \log(\frac{1+\sqrt{1-a^2}}{2}) & |a|>1.
    \end{dcases}
\end{align}
Combining everything yields
\begin{align}
    \label{eq:ReL_clean}
    \Re{L(xe^{i\theta})} &= \begin{dcases}
        \frac{1}{2}\log\left|\frac{J_R}{J_L}\right| &|x|<2\tilde{J}, \\
        \log(\frac{2|J_R|}{|x|+\sqrt{x^2-4\tilde{J}^2}}) & |x|>2\tilde{J},
    \end{dcases}
\end{align}
The maximum of the Lyapunov exponent thus happens at any point in the complex plane corresponding to the line segment of the OBC spectrum, depicted by the black line in \cref{fig:BBC} (a). On the other hand, the PBC spectrum is the curve whose values $E\in\mathds{C}$ are such that $\Re{L(E)}=0$ as the eigenstates are Bloch waves. 
\section{The Lyapunov Exponent for Rotationally-Invariant Eigenvalues}
\label{apdx:PointGap}

In this appendix, we assume that the OBC eigenvalues are rotationally-invariant about a point $E_c\in\mathds{C}$ and show various claims made in \cref{sec:Thouless}. Using the Thouless formula, we will prove that $\Re{L(E)}$ attains its maximum exactly at $\mathds{E}(V_n)$ and that 
\begin{align}
    \label{eq:apdx_continuation}
    \Re{L(E)} = \log(\frac{J_R}{|\mathds{E}(V_n)-E)|})
\end{align}
for all $E$ outside of the domain of OBC eigenvalues, where $J_R\equiv e^{\mathds{E}\log|J_{R,n}|}>0$ as in the main text.

First, we note that $\Re{L(E)}$ is a superharmonic function in the whole complex plane \cite{Craig1983}. This is easily seen from the fact that $\log |F|$ is subharmonic if the function $F$ is analytic, and that $F$ is subharmonic if and only if $-F$ is superharmonic. Since the set of superharmonic functions make a convex cone, then the expectation value of superharmonic functions is superharmonic. Thus, the real part of $L(E)$ satisfies the mean-value property: for any $E\in\mathds{C}$ and $r>0$, 
\begin{align}
    \label{eq:MVP}
    \Re{L(E)} \geq \int_0^{2\pi} \frac{d\theta}{2\pi}\,\Re{L(E+re^{i\theta})}.
\end{align}
Rotational invariance of the eigenvalues about $E_c$ implies that $\Re{L(E_c+re^{i\theta})}$ is independent of $\theta$ for any $r>0$, so \cref{eq:MVP} gives $\Re{L(E_c)}\geq \Re{L(E_c+re^{i\theta})}$ for any $\theta$. Since any $E\in\mathds{C}$ can be written as $E=E_c+re^{i\theta}$ for some $r>0$ and $\theta\in[0,2\pi)$, then
\begin{align}
    \Re{L(E_c)} &\geq  \Re{L(E_c +re^{i\theta})} =\Re{L(E)}.
\end{align}
Moreover, the mean of the energies $E_n$ of the OBC system is precisely $\mathds{E}(E_n)=E_c$. Because $\tr(H_0+V) = \tr(H_0) + \tr(V)$ for $H_0$ the clean Hamiltonian and $V$ the disorder, then $\mathds{E}(E_n) = \mathds{E}(V_n)$, where $E_n$ are the eigenenergies of $H_0+V$. This is true since the average energy of the clean system $H_0$ is zero. Thus, if the density of eigenvalues is rotationally-invariant about a point, then 
\begin{align}
    \Re{L(\mathds{E}(V_n))} \geq \Re{L(E)} \quad \forall E\in\mathds{C}.
\end{align}

We now show \cref{eq:apdx_continuation}. Suppose that all OBC eigenvalues are contained in the domain $ \{z\colon |z-\mathds{E}(V_n)|\leq R_0\}$ for some $R_0>0$, and let $E$ be outside of it. Using \cref{eq:sec2_Thouless}, we obtain
\begin{align}
    \begin{split}
    L(E) &= \log(\frac{J_R}{E-\mathds{E}(V_n)})\\
    &-\int d\rho(E_n)\,\log(1-\frac{E_n-\mathds{E}(V_n)}{E-\mathds{E}(V_n)}).
    \end{split}
\end{align}
where $\rho(E_n)$ is the probability distribution of OBC eigenvalues. Since $\rho$ is rotationally-invariant about $\mathds{E}(V_n)$, then we can write $\rho(E_n)=\rho_r(r_n)\rho_\theta(\theta_n)$, for $\rho_r$ and $\rho_\theta$ the radial and angular probability distributions, respectively. Moreover, $E_n-\mathds{E}(V_n) = r_ne^{i\theta_n}$ and by rotation invariance $\rho_\theta(\theta_n)=1/2\pi$ for all $\theta_n$. Thus,
\begin{align}
    \begin{split}
    &\int d\rho(E_n)\,\log(1-\frac{E_n-\mathds{E}(V_n)}{E-\mathds{E}(V_n)}) \\
    &= \int_0^{R_0}d\rho_r(r_n)\int_0^{2\pi}\frac{d\theta_n}{2\pi}\log(1-\frac{r_n}{E-\mathds{E}(V_n)}e^{i\theta_n}).
    \end{split}
\end{align}
Using the identity
\begin{align}
    \int_0^{2\pi} \frac{d\theta}{2\pi}\, \log(1-ze^{i\theta}) = \begin{dcases}
        0 & |z|<1, \\
        \log |z| &|z|>1
    \end{dcases}
\end{align}
with the fact that $r_n/|E-\mathds{E}(V_n)|<1$ for all $r_n\in[0,R_0]$, we find \cref{eq:apdx_continuation}.

\section{Correspondence Between $w(E)$ and $\Re{L(E)}$ in the Unidirectional Case}
\label{apdx:BBC_uni}

In this appendix, we prove \cref{eq:BBC} when $J_{L,n}=0$ for all $n=1,\dots,N$. As in \cref{sec:uni} of the main text, the OBC eigenenergies match the values of the on-site potentials $E=V_j$ and the associated eigenstates are
\begin{align*}
    \psi_n^{(j)}(E) &= \frac{1}{\mathcal{N}}\begin{cases}
        0 & 1\leq n < j, \\
        1 & n=j, \\
        \prod_{m=j+1}^n \frac{J_{R,m}}{E-V_m} & j<n\leq N
    \end{cases}
\end{align*}
for some normalizing constant $\mathcal{N}$. The Lyapunov exponent is then
\begin{align}
    L(E) &= \lim_{n\to\infty}\frac{1}{n}\sum_{m=j+1}^n \log(\frac{J_R}{E-V_m}).
\end{align}
Now, the quantity $\det (H_\text{pbc}^{(N)}(\Phi)-E)$ is calculated to be 
\begin{align}
    \det(H_\text{pbc}^{(N)}(\Phi)-E) &= \prod_{n=1}^N (V_n-E) - e^{-i\Phi}\prod_{n=1}^N(-J_{R,n}),
\end{align}
Taking the logarithm on both sides, we obtain
\begin{align}
    \begin{split}
    \log &\det(H_\text{pbc}^{(N)}(\Phi)-E)  = \sum_{n=1}^N\log(V_n-E) \\
    &\qquad\qquad+\log (1-(-1)^Ne^{-i\Phi}\prod_{n=1}^N\frac{J_{R,n}}{V_n-E}).
    \end{split}
\end{align}
In the large-$N$ limit, $e^{L(E)N}\sim\prod_{n=1}^N\frac{J_{R,n}}{V_n-E}$. Thus,
\begin{align}
    \begin{split}
    \log \det(H_\text{pbc}^{(N)}(\Phi)-E)  &= \sum_{n=1}^N\log(V_n-E)\\
    &\log (1-(-e)^{L(E)N}e^{-i\Phi})
    \end{split}
\end{align}
Taking the derivative with respect to $\Phi$ and integrating over $[0,2\pi]$ gives the spectral winding number $w(E)$ of \cref{eq:winding_num}. The argument principle gives $w(E) = Z-P$, where $Z$ and $P$ are the number of zeroes and poles of the function $f(z) = 1-(-e)^{L(E)N}/z$ inside the unit circle $|z|=1$ in the complex plane. There is obviously a pole at $z=0$, while the zero $|z| = e^{\Re{L(E)}N}$ is inside the unit circle when $\Re{L(E)}<0$ and outside otherwise. This establishes \cref{eq:BBC}.

\section{Periodic Potential Distributions}
\label{apdx:periodic_pot}
We here consider the topological phase transition between the phase of amplification and decay. We assume unidirectionality ($J_{L,n}=0$ for all $n=1,\dots,N$) and that the chain is made up of unit cells of length $L$. We note that this is an extension of the results given in \cref{sec:uni_SSPT}. Additionally, the questions we try to answer parallel those studied in Ref.~\cite{Fortin2025} with the bosonic Kitaev chain, but in the unidirectional model (in the notation of Ref.~\cite{Fortin2025}, with $t=\Delta$).

Suppose that the potentials $V_{nL+j}$ are distributed according to the probability distributions $f_j$ for $j=1,\dots,L$, where $f_j$ may depend on $j$. The $L$ dynamical equations are 
\begin{align}
    J_{R,nL+j}\psi_{nL+j-1} &= (E-V_{nL+j})\psi_{nL+j}
\end{align}
for $j=1,\dots,L$. Combining the $L$ equations together, 
\begin{align}
    \psi_{nL+j} &= \psi_{(n-1)L+j}\prod_{m=1}^L \frac{J_{R,(n-1)L+j+m}}{E-V_{(n-1)L+j+m}}
\end{align}
for all $j=1,\dots,L$. Therefore,
\begin{align}
    \psi_{nL+j} &= \psi_j \prod_{k=1}^n\prod_{m=1}^L  \frac{J_{R,(k-1)L+j+m}}{E-V_{(k-1)L+j+m}}.
\end{align}
Writing $V_n=r_ne^{i\phi_n}$, the condition $\Re{L(0)}=0$ for a topological phase transition in the large-$N$ limit is equivalent to 
\begin{align}
    \sum_{j=1}^L \int_0^\infty dr_n\, g_j(r_n)\log(\frac{J}{r_n}) &= 0.
\end{align}
where $J\equiv e^{\mathds{E}\log |J_{R,n}|}$. In particular, if the radial distributions $g_j$ are all uniform with support $[0,R_j]$, then the topological phase transition happens at $\sum_{j=1}^L\log(eJ/R_j) = 0$. On the other hand, if each potential $V_{nL+j} = V_j$ for $j=1,\dots,L$ and any $n$, that is, the system is $L$-fold translation-invariant, then the Thouless formula readily gives $w(V_j)=-1$: 
\begin{align}
    \Re{L(V_j)} &=   \frac{1}{L}\sum_{n=1}^L \log\frac{J}{|V_j-V_n|} = \infty.
\end{align}
Thus, the OBC eigenstates of energy $E=V_j$ are always skin states. This is readily generalizable to a discrete probabilistic distribution. If each potential $V_j$ in a discrete set $\{V_1,\dots,V_L\}$ has an associated probability $p_j>0$, then 
\begin{align}
    \Re{L(V_j)} = \sum_{n=1}^L p_n\log \frac{J}{|V_j-V_n|}=\infty
\end{align}
for all $j=1,\dots,L$. The OBC eigenstates are thus all skin states, while the PBC eigenstates are extended with $\Re{L(E)}=0$. As found in Ref.~\cite{Longhi2021}, when $\max_{1\leq j \leq L}|V_j|\ll J$, the mobility edge curve consists of a single loop. On the other hand, when $\min_{1\leq j \leq L} |V_j|\gg J$, the mobility edge curve splits into $L$ rings. The radii $\mathcal{R}_j$ of these rings are \cite{Longhi2021}
\begin{align}
    \mathcal{R}_j &= \left[\frac{J}{ \prod_{\substack{n=1 \\ n\neq j}}^L |V_n-V_j|^{p_n}}\right]^{1/p_j}.
\end{align}
The topological phase transition happens when $\Re{L(0)}=\sum_{n=1}^L p_n\log (J/|V_n|) =0$ and the winding number around the origin becomes zero when $\mathcal{R}_j<|V_j|$ for all $j=1,\dots,L$. 

\section{Extrema of the Lyapunov Exponent for a Dirac Delta Phase Distribution with Unidirectional Hopping}
\label{apdx:ReL_dirac_curve}
We here characterize the curve given by \cref{eq:ReL_dirac}. The first derivative in $x$ is 
\begin{align}
    \dv{}{x} \Re{L(xe^{i\theta})} &= \frac{1}{R_0}\log(\frac{R_0}{x}-1).
\end{align}
while the second derivative is 
\begin{align}
    \dv[2]{}{x}\Re{L(xe^{i\theta})} &= -\frac{1}{x(R_0-x)}<0.
\end{align}
The real part of the Lyapunov exponent is hence strictly concave on the line $E=xe^{i\theta}$ with $x\in[0,R_0]$. Setting the first derivative equal to zero gives that the maximum is attained at $x=R_0/2$, while the minima are attained at the endpoints $x=0,R_0$.

\bibliography{references.bib}

\end{document}